\documentstyle[12pt]{article}

\newcommand{\olV}{\overline{V}}
\vspace{-20ex}
\title{Ownership and Trade\\from Evolutionary Games}
\author{Kenton K. Yee${}^{\dag}$\\
{\sl Columbia University\/}}
\begin{document}
\bibliographystyle{bibstand}
\begin{titlepage}
\maketitle
\vspace{-42ex}
\mbox{\small \bf International Review of Law and Economics\/} \dotfill {\it forthcoming\/}
\vspace{40ex}

\begin{abstract}
\baselineskip=16pt
Ownership and trade emerge from anarchy as evolutionary 
stable strategies.  In 
these evolutionary game models, ownership status provides an endogenous 
asymmetrizing criterion enabling cheaper resolution of property conflicts. 
\end{abstract}
\vfill
\baselineskip=12pt
\noindent
${}^{\dag}$Assistant Professor, Columbia University,
615 Uris Hall, 3022 Broadway, New York, NY 10027.  The author
acknowledges financial support from
the Brown and Bain Fellowship in Law and High Technology,
the John M. Olin Fellowship in Law and Economics at Stanford
Law School, and the
Santa Fe Institute's Complex Systems Summer School.  

\begin{verbatim}
www.columbia.edu/~kky2001 
\end{verbatim}
\thispagestyle{empty}
\end{titlepage}
\pagebreak
\baselineskip=20pt
\parindent=10mm                

\baselineskip20pt
\setlength{\arraycolsep}{0pt}

\section{Introduction}
\label{sec:intro}

In the last three decades behavioral ecology 
has made strides in understanding
the origin and utility of behavioral traits
in animal societies (Maynard Smith 1988; Wilson 1975).  Much of
this success has come within the analytic framework of 
evolutionary game theory.  Evolutionary games, which 
mimic the dynamics of Darwinian natural selection, 
differ fundamentally from classical games (Maynard Smith 1982).  
While the latter 
focus on strategic interactions between rationally calculating
agents, evolutionary games model repeated interactions between 
adaptive but otherwise non-thinking agents.  

As one tends to believe humans think,
it is not surprising that
classical game theory has 
established a dominant position
in law and economics (Baird, Gertner, and Picker 1994)
while evolutionary game theory 
has attracted much less attention.\footnote{This is not
to suggest the absence of a strong evolutionary tradition
in law.  Scholars who have written about
the evolution of law include
Boyd and Richerson (1985); Cooter and Kornhauser (1980);
Elliot (1985); Ellickson (1991); Epstein (1980); Hirshleifer (1982);
Huang and Wu (1994);
Johnston (1996); Priest (1977); Rubin (1977); and Yee (1998).} 
A notable exception to this trend is
Hirshleifer, who pointed out
parallels between evolved animal behaviors (and their game theory models) and 
economically efficient human 
practices (Hirshleifer 1977, 1980, 1982).  Hirshleifer 
proposed
three metallic norms in particular: the ``Golden Rule of communal
sharing,'' the ``Silver Rule of private rights,'' and the ``Iron Rule
of dominance.''  Each, he asserted, has evolved
because they have sufficient socioeconomic advantages.  

This article develops an
evolutionary game model of property ownership and trade.  To start,
it extends and interprets
a well-known evolutionary model of animal territoriality
as a model of human property ownership.
That one could do this is perhaps not surprising.  The more surprising
result is that
trade -- the bilateral transfer of property for renumeration -- 
then emerges as a strategy which is evolutionarily preferred
over permanent ownership without trade.  In other words, traders are 
evolutionarily superior to owners who do not trade.  The 
identification that trade 
is an evolutionary stable strategy in evolutionary game theory
is the main contribution of this article.

The swallowtail butterfly, 
{\it Papilio\/} {\it zelicaon\/}, 
provides a provacative example of
what may underlie animal territorial behavior\footnote{
Other species, such as the hamadryas baboon, have been documented to 
exhibit similar behavior.} (Maynard Smith 1988, p. 214).  Because the
swallowtail lives in low density populations, 
one might expect matchmaking---the finding of a sexual partner---to 
be a problem.  This problem is solved in swallowtail society
by ``hilltopping.''
Males establish territories at or near the tops of hills and wait
for virgin
females, who instinctively seek out hilltops to mate.  Since there are
typically more males than hilltops, most males are relegated to
disadvantageous positions lower down
the slopes, where they attempt to waylay 
females on their way uphill.  Although the lower altitude males sometimes 
succeed, hilltop males mate most.  Curiously, despite their
enviable estates, hilltoppers are
seldom challenged by intruders.  On occasion when an intruder does 
confront a hilltopper, the visitor tends to
retreat after a brief, mild contest.

How is hilltop occupancy negotiated?  In 
experiments,  Larry Gilbert\footnote{Because a caterpillar virus
wiped out his butterfly population before Gilbert completed his
studies, Gilbert's original study, the one described in Maynard Smith 
(1988), was never published (L. E. Gilbert, private 
communication to Yee).  In any case,
R. Lederhouse describes similar observations of a closely related butterfly 
species (see Scriber, Tsubaki, and Lederhouse 1995).} tested two 
alternative explanations:
(A) Potential intruders are intimidated from invading because
hilltoppers are physically strongest, and because 
this fact is perceived by
the intruders;
or (B) Swallowtails defer 
to prior possessors, that is, whoever first stakes out a hilltop
is granted 
a socially-established
privilege to keep it (not unlike in the Anglo-American doctrine of 
adverse possession). 

In a series of experiments
with pairs of randomly selected male butterflies,
Gilbert ruled out A, and found support for the prior
possessor theory, B, as follows.  He convinced 
each butterfly of a pair that it was the sole occupier of 
the same hilltop by letting it be the hilltopper on 
alternate days.  On its days off, 
Gilbert kept the sidelined butterfly 
unaware of its counterpart's existence by confining it 
in dark room.  After a couple of weeks, each male clearly 
acted as
the hilltop's rightful proprietor, chasing away all 
comers, who invariably retreated without much protest.  When
Gilbert finally released both males to the same hilltop 
on the same day, an abnormally prolonged contest between the two 
``proprietors'' ensued, lasting many minutes 
and causing serious injury to each contestant.  As a result,
Gilbert concluded that deference to prior hilltop possessors
is an instinctive trait of swallowtails.  

Maynard Smith (1982) and 
others hypothesized that such instincts
evolved by natural selection of the fittest 
and constructed evolutionary game theories modeling
the evolutionary processes.  In these models,
deference to possession---a dispute resolution
strategy based on pre-existing status---is
evolutionarily preferred
over an always-fight strategy, which costs too much, and a never-fight
strategy, which yields too little.

Ownership conventions in human societies range from
the simplest unspoken norms, such as
not cutting ahead of somebody else in a grocery store queue
to much more involved rights bundles expressed in the Common Law.  A 
broad range of viewpoints (Rose 1985), ranging from
Locke's labor-mixing theory of property\footnote{
In a nutshell, Locke's view is that
one owns one's body and, by extension, 
the fruits of his body's labor.  Hence, first useful possession
establishes a property right (Epstein 1979).}
to law and economics\footnote{The normative
law and economics view
may be summarized as follows.  Property rights may be thought of
as a bundle of strands of primitive rights.  Particular
strands---tailored to the situation---are granted to
encourage social-welfare-optimizing use and investment.  A
benefit-cost balance determines
whether a particular strand should be granted
enforcement.  Costs include enforecement costs 
(Posner 1992).} may be called upon to justify
them.  Legal property rights can be
enforced by various combinations of 
liability and injunctive 
remedies (e.g. Calabresi and Melamed 1972).

Whatever the justifications, most
ownership rights bundles consist of two 
primitive strands: (a) Possession, 
the right to occupy or possess what one
owns, and 
(b) Trade, the right to buy and sell ownership.
By constructing evolutionary game theory models, I 
will illustrate how evolutionary 
forces can serve to establish these two strands as stable
strategies.  The trade model shows that those who trade are
evolutionarily preferred over Possessors who don't.  What 
is new here is that
{\it evolutionary forces are 
enough by themselves to establish Possession and,
given Possession, the practice of trade\/}.

Section~\ref{sec:ess} reviews Maynard Smith's (1982) construction of
evolutionary stable
strategy (ESS) and argues that ESSes have interpretation as
social norms.  Sections~\ref{sec:property} 
and \ref{sec:transfer} present 
two models corresponding, respectively, to
the two strands of ownership.  In the Section~\ref{sec:property} model, 
possession is an ESS; in Section~\ref{sec:transfer}, possession with 
the right to trade is an ESS.  

\section{Evolutionary Stable Strategies}
\label{sec:ess}

Social benefits
come only at a price: for every benefit accrued, there must be
set of behavioral constraints or
obligations to be fulfilled.  Social norms---prevalent
responses to recurring social situations---are 
the reciprocal constraints enabling the social benefits.   

Evolutionary game theory provides a quantitative 
dynamical theory of how 
such social constraints on behavior
can emerge from anarchy.\footnote{Axelrod (1986)
argues that social norms cannot survive unless there
is a secondary norm enforcing the first norm.  This implies the 
need for a tertiary norm to enforce the secondary
norm, and so forth {\it ad\/} {\it infinitem\/}.  (See
also Martinez Coll and Hirshleifer 1991; Lomborg 1996.)  
In contrast, the 
elegance of evolutionary stable strategies 
(as exemplified by the models herein)
is that, within the context of the game,
they do not require such an infinite hierarchy of
supporting norms.}
The basic ingredients of evolutionary game models for our purposes
are:
\begin{itemize}
\item a self-contained community of disputants
interacting in repeated, random pairwise encounters;
\item who, at each encounter, select from
a predetermined\footnote{In more ambitious formulations
the strategy space is permitted to evolve
via mutations in analogy to genes in biology.} menu of strategies, say,
$\{\alpha, \beta, \gamma, \cdots\}$ which
may be either pure or mixed.  
\item If player $\#1$ chooses strategy
$\alpha$ and $\#2$ chooses $\beta$, the payoff to player $\#1$ is
denoted $w_{\alpha\beta}$, which is determined 
entirely by $\alpha$ and $\beta$.
\item A round consists of many random encounters for each player.  After
each round,  
the community undergoes ``natural selection,'' in 
which strategies replicate
in proportion to how far above average their scores in the just-completed
round were; 
those with below average scores
die off in proportion to how far 
below average they were.
\end{itemize}

Consequently, an adaptive population evolves
according to a set of 
coupled first-order differential 
equations---one corresponding to each strategy---in the same spirit
as the Malthusian predator-prey relations.  Solving the equations yields
a phase diagram in strategy space, which typically has several fixed 
points (Friedman 1991).  Some of these fixed points are attractive:
populations always evolve into them so that they are 
evolutionarily stable against subversive mutations.  
Each attractive fixed point in a phase diagram represents an
``evolutionary stable strategy (ESS).''  

In other words, an ESS is a possible mode of
behavior that might lock-in.  Thus, an ESS can be interpreted as 
a social norm.  

For purposes of this article, it will be sufficient to state a criteria
defining ESS(es) given the game's defining payoff matrix.\footnote{From 
Maynard Smith (1982).  Other authors (e.g. Zeeman 1981) 
have pointed out that these criteria expressed in
terms of payoff matrix elements do not always recover all the
aforementioned attractive fixed points.  For our models, the technical
distinction is immaterial.}  To 
this end, for a given game strategy $\alpha_\star$ is an 
ESS if, starting from
a status quo where $\alpha_\star$ is the norm, it is not
possible for insurgents
to achieve higher payoffs with a renegade
strategy, say, $\gamma$.  Algebraically, this means $\alpha_\star$ is an ESS
if 
{\it either\/}
\begin{equation}
w_{\alpha_\star\alpha_\star} > w_{\gamma\alpha_\star} ~~ \forall \gamma
\label{ess1}
\end{equation}
{\it or\/}
\begin{equation}
w_{\alpha_\star\alpha_\star} = 
w_{\gamma\alpha_\star} ~ {\rm and\/} ~ w_{\alpha_\star\gamma} > 
w_{\gamma\gamma} ~~ \forall \gamma .
\label{ess2}
\end{equation}
Condition~(\ref{ess2}) is sufficient to establish an ESS because, 
even if
$w_{\gamma\alpha_\star} = w_{\alpha_\star\alpha_\star}$, 
invaders behaving according to 
$\gamma$ cannot successfully gain a foothold if  
they perform so poorly against each other that
they prevent themselves from becoming a sizable fraction of
the population.  Note that whether a strategy is an ESS
or not depends on the strategy space,
the set of competing strategies.  What is an ESS in one strategy
space may cease to be an ESS if the strategy space is expanded to
include other strategies.  Also, games
can (and usually do) have more than one ESS.

\begin{figure}
\[
\begin{array}{cccccc}
      &        &                   &{}_{\#2}&                   &        \\
      &        &  \quad H\quad     &        &  \quad D\quad     &        \\ 
\cline{3-5}
      &\vline{}&                   &\vline{}&                   & \vline{}\\
H\quad&\vline{}& \quad \Bigl( 
{V\over 2}-h, {V\over 2}-h \Bigr) \quad
&\vline{}& ~ \Bigl( V,0\Bigr) ~           & \vline{}\\
{}_{\#1}&\vline{}&                 &\vline{}&                   & \vline{}\\
\cline{3-5}
      &\vline{}&                   &\vline{}&                   & \vline{}\\
D\quad&\vline{}&  \Bigl( 0,V\Bigr) ~ & \vline{} &
~ \Bigl( {V\over 2}, {V\over 2} \Bigr) ~ & \vline{}\\
      &\vline{}&                   &\vline{}&                   & \vline{}\\
\cline{3-5}
\end{array}
\]
\caption{Payoffs to disputants $(\#1,\#2)$ in the two-player
``Hawk-Dove'' game.  Each disputant values the asset
at $\$V$.  Fighting has expected harm $\$h$ to each Hawk, and yields  
a ${1\over 2}$ chance of winning $\$V$.  If both are Doves (D),
they have a ${1\over 2}$ chance of getting the 
asset without fighting.  If one is a Hawk while the other is a Dove, 
the Dove retreats leaving the prize for the Hawk.
}
\label{fig:hawkdove}
\end{figure}

Of direct relevance to us is the Hawk-Dove game (Maynard Smith 1973),
whose payoffs are depicted in Figure~\ref{fig:hawkdove}.  In this game, 
two equally matched parties vie for the same asset,
worth $\$V$ to each.  (Imagine two randomly paired strangers 
fighting over a parking space at a crowded
shopping mall.)  Suppose only two strategies, Hawk (H) and Dove (D), 
are available.  Hawks always fight and, since both parties 
are equal, in a fight each
Hawk has only a one-half chance of winning the asset.
Fighting has an expected total 
cost to each participant of $\$h$.  In addition
to the expected injury, $h$ 
contains all other costs including the expected energy expenditure
and any risk-bearing costs.  Doves 
retreat when confronted by a Hawk.  If two Doves
meet, a random one of the two Doves retreats 
and leaves the other to the spoils.

In neoclassical game theory (e.g. 
Baird, Gertner, and Picker 1994),
${1\over 2} V>h$ corresponds to the Prisoner's Dilemma while
${1\over 2} V<h$ corresponds to the Chicken game.
In the Prisoner's Dilemma, $H$ is called ``defection'' and defection is
the unique Nash equilibrium strategy for both players.  In the Chicken game,
the unique Nash equilibrium is for one player to be the chicken (D)
and the other to be the hawk (H).

When ${1\over 2} V > h$ (the Prisoner's Dilemma case),  
H is an ESS and D is {\it not\/} by virtue of
Criterion~(\ref{ess1}).  In other words, aggressiveness
is evolutionarily preferred when the rewards outweigh the costs
of fighting.  Note that this does {\it not\/} mean
H optimizes social welfare.  In fact, an all-Dove population 
maximizes social welfare.

When the possibility of
serious injury is sufficiently large,
${1\over 2}V < h$ (the Chicken game case).  In this case,
neither pure strategies $H$ or $D$ are ESSes.  It turns out
that the only ESS (within the H-D strategy subspace) 
is a mixed strategy
$\alpha_\star = 
 p H + (1-p) D$, where 
$p \equiv {1\over 2}V/h$. 
This mixed strategy can be achieved in two ways.  Either the population
is homogeneous and at an encounter
each individual exercises H with probability $p$, or
the population is comprised of fraction $p$ Hawks and $(1-p)$ Doves.
In either case, the more potential costs exceed potential gains
the less evolutionarily attractive 
hawkishness is.

The material in this section was established by 
Maynard Smith (1982) and others.  While the Hawk-Dove game  
has provided insights into animal behavior,
it is too simple to allow for more sophisticated
human strategies.  Humans act
like neither Hawks, Doves, nor mixtures thereof.  Rather, 
we have more sophisticated options.  The next section turns to 
one of them.



\section{Possession as an ESS}
\label{sec:property}

\begin{figure}
\[
\begin{array}{cccccccc}
      &        &                   &        &\#2      &        &    &        \\
      &        &  \quad H\quad     &        &  \quad\quad D\quad\quad    &        &  
\quad P\quad  &   \\ 
\cline{3-7}
      &\vline{}&                   &\vline{}&              &\vline{}&    &\vline{}\\
\quad\quad~H\quad&\vline{}&
~\Bigl( {V\over 2}-h, {V\over 2}-h \Bigr)~ &\vline{}& \Bigl(V,0\Bigr) 
  &\vline{}& ~\Biggl(
{ ({V\over 2}+h) f  \atop ~ + {V\over 2} -h }, 
( {V\over 2} -h ) f \Biggr) ~ &\vline{}\\
      &\vline{}&                   &\vline{}&              &\vline{}&    &\vline{}\\
\cline{3-7}
      &\vline{}&                   &\vline{}&              &\vline{}&    &\vline{}\\
\#1\quad~D\quad&\vline{}& 
~\Bigl( 0, V\Bigr) ~&\vline{}&\Bigl( {V\over 2}, {V\over 2}\Bigr) &\vline{}&
	  \Bigl( {V\over 2} f, {V\over 2}(1+f) \Bigr)    &\vline{}\\
      &\vline{}&                   &\vline{}&              &\vline{}&    &\vline{}\\
\cline{3-7}
      &\vline{}&                   &\vline{}&              &\vline{}&    &\vline{}\\
\quad\quad~P\quad &\vline{}& ~\Biggl( \bigl({V\over 2}-h\bigr) f, 
{\bigl({V\over 2}+h\bigr)f \atop ~ + {V\over 2}-h } \Biggr) ~ &\vline{}& 
~\Bigl( {V\over 2} (1+ f ), {V\over 2}f\Bigr)~ &\vline{}&
						     \Bigl(f  V, fV\Bigr) 
&\vline{}\\
      &\vline{}&                   &\vline{}&              &\vline{}&    &\vline{}\\
\cline{3-7}
\end{array}
\]
\caption{Payoffs to disputants $(\#1,\#2)$ in the 
``Hawk-Dove-Possessor'' game.  If Possessor P owns
a piece of land, it will fight for it as a Hawk; if P's 
opponent is the owner, P defers to him and acts as a Dove.   
$f$ is the fraction of time on average a disputant expects to
be in the role of owner.
}
\label{fig:hdp}
\end{figure}

``Finder's keepers''
and ``first come, first serve'' are not only basic thumb rules
in playground citizenship,
they are powerful norms that have been recognized by the courts
and applied widely in such varied settings as 
adverse possession,
abandoned property, fisheries, wildlife, seabed minerals,
groundwater rights, intellectual property, debt collection, 
oil and gas, pollution permits, the radio frequency 
spectrum, satellite orbits, and ownership of wartime spoils.

To model this norm as an ESS, introduce
the ``Possessor (P)'' 
strategy: 
\[
P \equiv \cases{{\rm H\/} & if current owner;\cr 
		{\rm D\/} & if current intruder.}
\]
The Possessor strategy models the practice
of ``possession''.  Unlike Hawks and Doves, 
Possessors observe convention based on their status; their
behavior depends on whether they are the owner or intruder.

Figure~\ref{fig:hdp} depicts the
Hawk-Dove-Possessor game payoffs.  If 
$f={1\over 2}$, the Hawk-Dove-Possessor game reduces to 
the Hawk-Dove-Bourgeois
game (Maynard Smith 1988; Hirshleifer 1982, pp. 22-23).  What 
distinguishes the Hawk-Dove-Possessor
game from the Bourgeois game is the parameter $f$.  $f$ is the
expected fraction of confrontations in which
a disputant anticipates she will be in the role of Owner.  We 
will show that P is the unique ESS
for \underline{all} 
$f$ in the interval $(0,1)$.  This 
implies that the Possessorship strategy is robust to 
the wealth (or dearth) of ownership opportunities available -- all one
requires is a nonzero chance ($f\ne 0$) to be an owner.

The payoffs of the Hawk-Dove-Possessor game are motivated as follows.  
First, the parameter $f$ has the following interpretation.  Imagine 
a neighborhood in which everyone possesses
a separate (but otherwise identical) 
plot of land.  At the same time, everyone is wandering
the neighborhood interested in obtaining possession of additional
plots.  Suppose that each agent
wanders randomly throughout the neighborhood so
that, in the repeated
disputes which arise, each disputant is in the role of 
owner fraction $f$ the time and intruder
the other $1-f$.  For instance, if
there were $N$ disputants each owning one plot of land, 
in a round-robin tournament every disputant would intrude
$N-1$ times and be reciprocally intruded upon $N-1$ times.  In 
this case, $f={1\over 2}$ because
every disputant is an owner half the time and an intruder the other half.

Also of interest is when
land is scarce and not everyone can own a plot.  Then
the probability $f$ 
of being an owner in a random encounter is no longer ${1\over 2}$, but
will be some positive number less than that.
To obtain an expression for $ f $, 
let $n$ denote the number of available plots, and $N$ the
total number of disputants.  Assume $N\ge n$ and that $N$ is very large.  
Assume everyone has an equal chance to be an owner
and nobody owns more than one plot
of land at any time.  Then the chance to be an owner is
$g = {n\over N}$.  The 
total number of encounters a plot-owner endures in a 
round-robin tournament where every disputant (owners and non-owners alike)
attempts to intrude once on every alien plot is
$(N-1)+(n-1)$ because the owner defends his land once against the other
$N-1$ disputants and intrudes once on the other $n-1$ plot owners.  
Hence, ignoring the higher order effect of what happens
when property transfers resulting from encounters lead to a disputant
temporarily owning more than a single plot of land, the fraction of
encounters where {\it an owner\/} plays the role of owner is
\[
{N - 1 \over N-1 + n-1} ~.
\]
The overall fraction
$f$ of time a random disputant plays the role of owner in an encounter is
\begin{eqnarray}
 f  &~=~ & 
\Bigl( {\rm chance~to~be~an~owner\/}\Bigr) ~\times ~
\Biggl( {{\rm fraction~of~encounters\/} \atop 
{\rm in~owner's~role~if~owner\/}} \Biggr)
\nonumber\cr
  & ~=~ & g \times \Biggl({N-1 \over N-1 + n-1}\Biggr)
\nonumber\cr
  & ~=~ & g \times \Biggl({1-{1\over N} \over 1 + {n\over N}-{2 \over N}} 
\Biggr) ~.
\nonumber
\end{eqnarray}
In the limit where the number $N$ 
of disputants is large and the ratio $g=n/N$ is finite ($n$ may or may
not be large depending on the value of $g$, but $n$ is strictly
less than $N$),
\[
\lim_{N\mapsto\infty\atop
g=n/N} f \sim {g\over 1+ g} .
\]
In this limit, $0\le  f  \le {1\over 2}$ because $0 < g < 1$.  In 
particular, $f = 0$ when the chance of land ownership is 
zero ($g=0$).  When land is scarce ($n << N$), the chance
of land ownership is small ($g << 1$) and so is $f\sim g$.  On 
the other hand, when the chance of land ownership is
almost certain\footnote{When land is plentiful relative to
the number of disputants ($n>> N$), the chance $g$ for a disputant
to be an owner is given by a complex combinatoric function if
disputants are allowed to own more than one plot of land; 
the expression $g=n/N$ is valid only when ownership is capped
at one plot per person.}
($g\mapsto 1$), then $f  = {1\over 2}$. 

In the Hawk-Dove-Possessor game, at each encounter a disputant can 
either behave like a Hawk, a Dove, or
a Possessor.  Figure~\ref{fig:hdp} depicts the payoffs of
the Hawk-Dove-Possessor game.  In deriving Figure~\ref{fig:hdp}, 
payoffs have been averaged over many encounters.  Since disputants are
owners $f$ of the time, and intruders $1-f$
of the time, the payoff to disputant $\#1$ is the
weighted average of two conditional payoffs:
\[
w =  f  w_{\vert~ {\rm owner\/}}
+ (1- f )  w_{\vert~ {\rm intruder\/}} .
\]
For example,
\[ 
w_{PH} =  f  ({1\over 2}V-h) +
(1- f ) \times 0  = ({1\over 2}V-h) f, 
\]
\[ 
w_{HP} =  f  V + (1-f) ({1\over 2}V-h) = 
{1\over 2}V-h + ({1\over 2}V+h) f, 
\]
and
\[
w_{PP} =  f  V  + (1- f ) \times 0 =  f  V.
\]

When ${1\over 2} V < h$ (the Chicken game case), P is the only 
ESS of the Hawk-Dove-Possessor game.  (The reader is invited to
confirm this using Conditions~(\ref{ess1}) and~(\ref{ess2}).) In fact,
one can go further and prove there is no mixed ESS so that P is the
unique ESS if the strategy space is H-D-P.)  As this is true 
for {\it any\/} value of $ f \in (0,1)$, it takes only 
a {\it tiny\/} (any nonzero) chance $f$ of ownership
to establish P as an ESS.  The fact that P is not an ESS exactly at
$ f  = 0$ is moot because $ f  = 0$ means there is
no property to be owned ($n=0$).

In defining the 
Hawk-Dove-Possessor payoff matrix, I assumed that owners and
intruders value the disputed plots equally, that it costs
owners and intruders the same to engage,
and that owners and intruders each have equal chances 
of winning a fight.  In real life, due to informational advantages of
being an owner, owners probably value and can defend their
properties more and better than intruders.  So it is
likely that possessorship is even more preferred than in this stylized
model.  What the Hawk-Dove-Possessor
model shows is that 
{\it possessorship\/} {\it is\/} {\it evolutionarily\/} {\it stable\/}
{\it despite\/} {\it ignoring\/} {\it all\/} {\it the\/}
{\it likely\/}
{\it advantages\/} {\it an\/} {\it owner\/} {\it has\/} {\it over\/}
{\it an\/} {\it alien\/} {\it intruder\/}. 
 
While P is an ESS for the Chicken game, it does
not resolve the Prisoner's Dilemma.  When 
${1\over 2} V > h$ (the Prisoner's Dilemma case), 
$H$ remains the only pure-strategy
ESS.  When the expected gain 
from fighting exceeds expected losses, aggressiveness is evolutionarily 
preferred over dovishness and possessorship.  

Possessor
is an ESS for the Chicken game because it
provides a predictable and costless mechanism for resolving
disputes.  As in the game, possession in the real world serves to
establish an asymmetric dispute resolution mechanism in an
otherwise symmetric situation.  This role is consistent with
the prolific public notice requirements
usually associated with possession, including
title registration requirements and the
adverse possession doctrine, 
under which owners can be divested of property for failing to protest
against dispossession in a timely manner.  

The utilitarian 
value of possession was recognized early on in the
common law.  In {\sl Pierson v. Post\/}\footnote{
\underline{Pierson v. Post\/}. 3 Cai. R. 175 (N.Y. Sup. Ct. 1805)}
a famous wild-fox case from the nineteenth century, a hunter, Post, 
had a fox in his gunsight but before he could fire 
an interloper killed the fox
and ran off with the prize.  The indignant Post sued the interloper
on the theory that his pursuit of the fox established
his right to have it.  The court disagreed.  It held that
possession requires a clear act putting the world
on notice that the ``pursuer has an unequivocal intention 
of appropriating the animal to his individual use.'' (Id. at 178.)  
Gaining property rights over a wild animal requires
either establishing physical control over it or mortally wounding it.

\subsection{Why not Anti-Possessor?}

I asserted at the beginning of Section~\ref{sec:property}
that the virtue of Possessorship 
is dispute resolution via an objective symmetry-breaking criterion.
Yet, Possessorship is not the only symmetry-breaking
mechanism one can imagine.  In particular, consider its mirror image, 
the ``Anti-Possessor (AP) strategy:''
\[
AP \equiv \cases{{\rm D\/} & if current owner;\cr 
		{\rm H\/} & if current intruder.}
\]

One might reasonably 
guess that Anti-Possessorship would serve just as well as
Possessorship as a symmetry-breaking device.   AP indeed 
{\it is\/} an ESS in a game consisting of Hawks, Doves, and Anti-possessors.

In animal societies, AP is
rare, but not totally unheard of.  The social spider
{\it Oecibus\/} {\it civitas\/} lives together in groups, but
each constructs its individual web.  If a spider is driven from
its web, it may dart into the web of a neighbor.  If the neighbor
is in residence, it does not expel the intruder, but instead darts
out into somebody else's web.  As a result, dislocating a single spider may
trigger a chain reaction---a game 
of musical webs (Maynard Smith 1982, p. 96).

It is apparent why P is more common than AP in both 
nature and culture.  AP is disfavored due to 
nomadicy costs.  In an AP culture, citizens are forced to
alternate as owners and intruders in an unending tag-team match.  Due
to relocation costs, AP is not as viable as P.

\section{Trade as an ESS}
\label{sec:transfer}

While possession retention is the first strand of property rights, 
the trading is the second strand.  Trade is efficient because if 
an intruder values a property at $\$V$ while
the current owner values it at $\$v < \$V$, then
both benefit if the owner sells it to the intruder 
for $\$x$ where 
\[
v < x < V .  
\]
Beyond restricting $x$ to the interval $(v,V)$, the model here
does not have anything to say about whether the transaction
price $\$x$ is paid in money or
another asset of value $\$x$.  The model also does not
specify how the value of $x$ is determined -- whether
by case-by-case negotiation or by an exogenously
given recipe.  In this section, I will simply show that 
trading is an ESS for {\it any\/} 
$x \in (v,V)$.  The point is that trade is evolutionarily 
preferred over 
stoic possessorship (P). 

Suppose in every encounter, the two 
disputants do not value the disputed 
property equally because, for example, they always have slightly
different reasons for wanting the property.
Let $\$V$ and $\$v$, with $v < V$, denote the two mismatched valuations.
Define a ``Trader (T)'' as a Possessor who 
is willing to sell or buy for $\$x$ when dealing with a fellow Trader.
In particular, when both owner and
intruder of a particular encounter 
are Traders, and the intruder values the property more (e.g., 
at $\$V$) than the owner, 
he will want to purchase the property from the 
owner for $\$x$, where $v < x < V$.   Symbiotically, a Trader-owner
who values the property less (e.g., at $\$v$) than a Trader-intruder
will readily agree to sell the property to the latter
for $\$x$.  In other words,
\[
T \equiv \cases{ ~ ~ ~ {\rm ~ ~ ~  P\/} & if counterpart is not T;\cr
                \cases{{\rm sell~for\/}~x & if $v$-valuing owner;\cr
                       {\rm buy~for\/}~x & if $V$-valuing intruder;\cr
		       ~ {\rm ~~~ P\/}  & otherwise;   }& if counterpart is T.}
\]

\begin{figure}
\[
\begin{array}{cccccccccc}
      &        &          &         &        &{}^{\#2}   &    &    &    &        \\
      &        &  \quad H\quad     &        &  \quad D\quad    &        &  
\quad P\quad  &  & \quad T\quad &   \\ 
\cline{3-9}
      &\vline{}&                   &\vline{}&              
&\vline{}&    &\vline{}  & & \vline{} \\
\quad\quad~H\quad&\vline{}& 
~\bigl( w_{HH},w_{HH}\bigr)~ &\vline{}& ~\bigl(w_{HD},w_{DH}\bigr) ~ 
&\vline{}& ~\bigl( w_{HP}, w_{PH} \bigr) ~
& \vline{}& ~\bigl( w_{HT}, w_{TH} \bigr) ~ &\vline{}\\
      &\vline{}&                   &\vline{}&              
&\vline{}&    &\vline{}  & & \vline{} \\
\cline{3-9}
      &\vline{}&                   &\vline{}&              
&\vline{}&    &\vline{}  & & \vline{} \\
{}_{\#1}\quad~D\quad&\vline{}& 
~\bigl( w_{DH},w_{HD}\bigr)~ &\vline{}& ~\bigl(w_{DD},w_{DD}\bigr) ~ 
&\vline{}& ~\bigl( w_{DP}, w_{PD} \bigr) ~
& \vline{}& ~\bigl( w_{DT}, w_{TD} \bigr) ~ &\vline{}\\
      &\vline{}&                   &\vline{}&              
&\vline{}&    &\vline{}  & & \vline{} \\
\cline{3-9}
      &\vline{}&                   &\vline{}&              
&\vline{}&    &\vline{}  & & \vline{} \\
\quad\quad~P\quad &\vline{}& 
~\bigl( w_{PH},w_{HP}\bigr)~ &\vline{}& ~\bigl(w_{PD},w_{DP}\bigr) ~ 
&\vline{}& ~\bigl( w_{PP}, w_{PP} \bigr) ~
& \vline{}& ~\bigl( w_{PT}, w_{TP} \bigr) ~ &\vline{}\\
      &\vline{}&                   &\vline{}&              
&\vline{}&    &\vline{}  & & \vline{} \\
\cline{3-9}
      &\vline{}&                   &\vline{}&              
&\vline{}&    &\vline{}  & & \vline{} \\
\quad\quad~T\quad &\vline{}&  
~\bigl( w_{TH},w_{HT}\bigr)~ &\vline{}& ~\bigl(w_{TD},w_{DT}\bigr) ~ 
&\vline{}& ~\bigl( w_{TP}, w_{PT} \bigr) ~
& \vline{}& ~\bigl( w_{TT}, w_{TT} \bigr) ~ &\vline{}\\
      &\vline{}&                   &\vline{}&              
&\vline{}&    &\vline{}  & & \vline{} \\
\cline{3-9}
\end{array}
\]
\[
\begin{array}{ccccccccccc}
      & &        &          &         &        &  &    &    &    &        \\
      & &        &  \quad H\quad     &        &  \quad D\quad    &        &  
\quad P\quad  &  & \quad T\quad &   \\ 
\cline{4-10}
    &  &\vline{}&                   &\vline{}&              
&\vline{}&    &\vline{}  & & \vline{} \\
& \quad\quad~H\quad&\vline{}&{V+v\over 4}-h &\vline{}& {V+v\over 2}
   &\vline{}& ~ {{V+v\over 4}(1+ f )\atop ~~~ - h (1- f )} ~& 
   \vline{}& {{V+v\over 4}(1+ f )\atop ~~~ - h (1- f )} &\vline{}\\
 &     &\vline{}&                   &\vline{}&              
&\vline{}&    &\vline{}  & & \vline{} \\
\cline{4-10}
   &   &\vline{}&                   &\vline{}&              
&\vline{}&    &\vline{}  & & \vline{} \\
w_{\alpha\beta}
~\equiv ~ & \quad\quad~D\quad&\vline{}& 0&\vline{}&{V+v\over 4} &\vline{}&
	  		{V+v\over 4} f   & \vline{}
                   &    {V+v\over 4} f   &\vline{}\\
&      &\vline{}&                   &\vline{}&              
&\vline{}&    &\vline{}  & & \vline{} \\
\cline{4-10}
   &   &\vline{}&                   &\vline{}&              
&\vline{}&    &\vline{}  & & \vline{} \\
& \quad\quad~P\quad &\vline{}& ~{V+v\over 4} f  - h f ~  
  &\vline{}& ~{V+v\over 4}(1+ f ) ~ &\vline{}& 
             {V+v\over 2} f  &\vline{}& {V+v\over 2} f  
             &\vline{}\\
 &     &\vline{}&                   &\vline{}&              
&\vline{}&    &\vline{}  & & \vline{} \\
\cline{4-10}
 &     &\vline{}&                   &\vline{}&              
&\vline{}&    &\vline{}  & & \vline{} \\
 & \quad\quad~T\quad &\vline{}&  ~{V+v\over 4} f  - h f ~  
  &\vline{}& ~{V+v\over 4}(1+ f ) ~ &\vline{}&
	{V+v\over 2} f   &\vline{}& ~{V\over 2} - ({1\over 2}- f )x~ 
  &\vline{}\\
      & &\vline{}&                   &\vline{}&              
&\vline{}&    &\vline{}  & & \vline{} \\
\cline{4-10}
\end{array}
\]
\caption{The top table depicts the payoffs to disputants $(\#1,\#2)$ 
in the Hawk-Dove-Possessor-Trader game.  The lower table 
lists the payoff values $w_{\alpha\beta}$
for $i,j\in \{H,D,P,T\}$.  A disputant is an owner
fraction $f$ of the time and an intruder $1- f$ of the time.
One party values the disputed
property at $\$V$ and the other at $\$v < \$V$.  Half the time, 
the $V$-valuer is the owner, half the time
the intruder.  If both owner and
intruder are Traders (T) and the intruder values the property more, 
he purchases from the owner for $\$x$ where $v < x < V$. 
}
\label{fig:hdpt}
\end{figure}

Figure~\ref{fig:hdpt} depicts the payoffs of the Hawk-Dove-Possessor-Trader
game (``Trader game'' for short).  In 
deriving the payoff matrix, I have assumed that
disputants are owners fraction $ f $ of the time, and intruders $1- f $
of the time.  Additionally, each encounter
is between a disputant valuing the property at $\$V$ and one
valuing it at $\$v$.  Thus, the payoff to disputant $\#1$ is the
weighted average of four conditional payoffs:
\[
w = { f  \over 2} (
w_{\vert~ {V-{\rm valuing\/} \atop {\rm owner\/}}} 
+ 
w_{\vert~ {v-{\rm valuing\/}\atop {\rm owner\/}}} ) 
+ {1- f \over 2} ( 
w_{\vert~ {V-{\rm valuing\/}\atop {\rm intruder\/}}} 
+ 
w_{\vert~ {v-{\rm valuing\/}\atop {\rm intruder\/}}} ) .
\]
For instance,
\[
w_{TT} = { f \over 2} \bigl( V + x ) + {1- f \over 2}
( V-x + 0 ) = {1\over 2} V - ({1\over 2}- f ) x .
\]

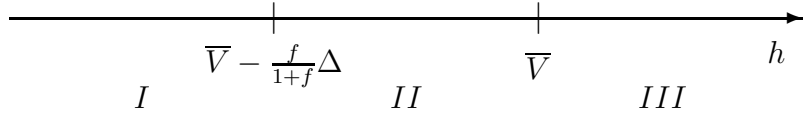
\begin{figure}[tbp]
\begin{picture}(350,65)(-50,0)
\thicklines
\put(100,63){\makebox(0,0){$|$}}
\put(100,45){\makebox(0,0){$\olV - {f\over 1+f} \Delta$}}
\put(200,63){\makebox(0,0){$|$}}
\put(200,45){\makebox(0,0){$\olV$}}
\put(50,33){\makebox(0,0){$I$}}
\put(150,33){\makebox(0,0){$II$}}
\put(247,33){\makebox(0,0){$III$}}
\put(290,50){\makebox(0,0){$h$}}
\put(0,63){\vector(1,0){300}}
\end{picture}
\caption{$\olV \equiv {V+v\over 4}$ is average value
each party would walk away with if they equally shared the asset.
$\Delta \equiv {V-v\over 2} > 0$ is 
the {\it incremental\/} expected gain of trading.  $h$ is the
cost of fighting.  
In Region I (where $h < \olV - {f\over 1+f}\Delta$), H is the
unique pure strategy ESS.  
In Region II (where $\olV- {f\over 1+f}\Delta < h< \olV$), 
both T and H are possible ESSes.  In Region III 
(where $h > \olV$), T is the unique pure strategy ESS.} 
\label{fig:regions}
\end{figure}

The ESSes of the Trader game depends
the relative values of
the cost of fighting $h$, the average spoils 
of sharing without fighting $\olV \equiv {V+v\over 4}$, and
the {\it incremental\/} expected per party gain from trading 
${V-v\over 2}$.  As depicted in Figure~\ref{fig:regions}, when
the cost of fighting is small (Region I), H is an ESS and T
is not.  On the other hand, when the cost of 
fighting is comparably large (Region III), T is an ESS
and H is not.  

When fighting cost is intermediate (Region II), both H and T are ESSes.  
If $V=v$, then $\Delta = 0$ and Region II is subsumed by
Region I.  Hence, Region II owes its existence to a valuation discrepancy
between owner and intruder.  In Region II, hawkishness is viable 
because the cost of fighting is less than the expected spoils $\olV$.  
At the same time, the cost of fighting exceeds the 
expected harm from not Trading, so Trading is also viable.
As it is, if everyone is trading in Region II, a disputant is better
off to trade.  On the other hand, if everyone is a Hawk
in Region II, a disputant is advised to also be a Hawk.

These results follow from
verifying the necessary and sufficient ESS 
Criteria~(\ref{ess1}) and (\ref{ess2}).  The intuition for
why Region III ($h > \olV $) is a
{\it trade-only\/} region is as follows.  In this case, within the H-D-P 
strategy subspace
P is the ESS (for the same reasons why P is the ESS in the
Hawk-Dove-Possessor game).  T fares no worst than P against Hawks and
Doves since Traders and Possessors act the same way against Hawks
and Doves.  The remaining issue is why T fares better than P within
the P-T subspace.  The reason is because, in trades 
the seller-owner always 
nets a surplus of $\$(x-v)$ while the
buyer-intruder nets a surplus of $\$(V-x)$.  Because both parties gain
and nobody loses in a trade, Traders fare better than
Possessors and T is the only pure-strategy ESS.

Clearly, Region III corresponds to bartering in ancient societies
and commerce in modern ones.  When the cost of fighting is
high, trade evolves. 

Likewise, Region I is reminiscent of 
nations fighting over
land, oil fields, and governance structures.  When the expected value of
the asset far exceeds the cost of war, hawkishness emerges.

Region II, when fighting cost is intermediate, 
is the most interesting.  In this case, {\it either\/}
H or T can emerge and, once one of them does, it locks in
as the ESS.  However, either H or T has an equal {\it ex ante\/}
chance of emerging.  For instance, Region II may help explain 
some litigation practices in the
United States, but whether it does or not is
difficult to assess because determining the parties' costs and
benefits in these situations is difficult.  Suppose
costs are such that Region II applies in some legal disputes,
such as bankruptcy or tort cases.  Then bankruptcy cases
may be more likely settled out of court (a T strategy)
while tort litigation may go to trial
more frequently (an H strategy) simply because
historically a T strategy has emerged
as normal practice in bankruptcy while an H strategy is the norm
in tort litigation.  Once an ESS is established, 
the strategy (T or H) becomes the {\it modus operandi\/} when
that kind of dispute arises
even though the other strategy could fare equally well if everyone
were to adopt it instead.

Finally, Figure~\ref{fig:regions} highlights 
the cost $h$, which is exogenously given,
as the critical determinant of whether
T or H or both are evolutionarily favored.  As law has the
power to change $h$ by exacting extra penalties and fees
on disputants, an implication 
of the Trading game is that law can inspire or hinder
the evolution of trade by its assessment of penalties on
hawkishness or trading.

\section{Concluding Remarks}

The Hawk-Dove-Possessor game and the Hawk-Dove-Possessor-Trade game 
provide two messages.  The first message is that
deference by intruders to owners is evolutionarily preferred over
non-status-based behavior.  This is because prior possession provides
a ready asymmetrizing criterion enabling resolution of 
possession conflicts.  Individuals who avoid transactions costs
by resolving conflicts based on a cultural asymmetrizing criterion
are better off than those who do not.

Can one do better than mere possession?  The second message 
answers this question affirmatively.  A Trade strategy trumps 
Possessorship with no trade.  Trade is 
the ability to buy and sell according to
what optimizes personal gain; trading does not occur unless both parties
gain.  Accordingly, traders always benefit from trade and, so, are
evolutionarily preferred.  Analysis of the 
Hawk-Dove-Possessor-Trade game shows that those who trade are
evolutionarily preferred over Possessors who don't.  That
voluntary exchange tends to improve social welfare is not new; 
it is already well
known from traditional Walrasian analysis.  What is new here is that,
{\it without the help of rationality or utility maximization, evolution 
is enough by itself to generate trade\/}.

This raises the question, If trade is evolutionarily preferred over 
possession without trade, why don't Gilbert's butterflies in
Section~\ref{sec:intro}
trade?\footnote{To be more precise, there is no evidence to
suggest that butterflies trade.  However,
we cannot rule out the possibility that they do trade.  Other
species exhibit a rudimentary form of trade biologists call
``mutualism.''  Cleaning fish, which eat debris off the bodies of bigger
fish, trade their cleaning services for food (Chapter 13,
Maynard Smith 1982).}  Why doesn't an animal 
with extra food barter away some in exchange for
a future meal?  To be sure, this does occur in a rudimentary way 
when animal family groups hunt together in cooperative 
packs and share prey.  But 
it is probably fair to say that animals do not trade with strangers 
unless the exchange is immediately mutually gratifying (as in the
cleaning fish example).  

The main barrier to animal trade is probably logistical and traces
back to the exchange value $x$ in the Hawk-Dove-Possessor-Trade game.  The
existence of $x$ tacitly presumes a form of book keeping, either
mental accounting or the use of money.  Money has not evolved in
the animal kingdom and, 
absent money, anonymous trading is difficult because
unrelated animals cannot keep track of who owes who how much.  Hence, 
the failure of animals to evolve a monetary system
probably has obstructed the emergence of animal
trade from animal territoriality instincts.  But why has
money not evolved in animal kingdoms?  Perhaps, like 
law or other sociopolitical structures, money is higher up
in the evolutionary path, and it emerges only along with human (or
prehistoric human) traits such as language and agriculture.  

Beyond property rights, can evolutionary games
explain the origin of 
other human social norms?  An affirmative answer requires two 
ingredients: a theory
of the origin of social norms, and an identification of the 
specific evolutionary
games corresponding to each norm.  This article has illustrated
how to realize the second of these two ingredients by constructing
the games corresponding to the practice of Possession and Trade.  Let's 
now turn to the first.

For an evolutionary theory of social norms, it is unlikely that
human social norms are biologically evolved
like butterfly territoriality is.  Rather, 
the allusion to biological evolution must be metaphorical.  In 
biological evolution, genes undergo natural selection in
an ecology of water, geology, food, and weather.  To survive,
genes and their organic manifestations must adapt,
mutate, and establish parasitic, predatory, or mutualistic niches in
the food chain.  Natural selection of the fittest
weeds out inappropriate genes.  In the long run
only genes which manage to stake out biological niches in
the food chain reproduce and propagate.  

Norms are the ``genes'' undergoing natural selection.
Through their embodiment in the behaviors of their human
carriers, norms compete against each other for social popularity
against a backdrop of cultural traditions and
legal and political institutions -- themselves manifestations of
successful ideas.  Not unlike between genes in biological evolution,
the competition between norms for social acceptance
and influence is a life and death struggle.  Those
that don't successfully adapt across time and sociopolitical barriers die.

This notion that norms are like genes is not new.
In \underline{The\/} \underline{Selfish\/}
\underline{Gene\/}, Dawkins (1976)
proposes that social ideas, what he calls ``memes,''
are a nonorganic form of life.  His 
examples of memes include tunes, catch-phrases,
taboos, and architectural fashions.  In Dawkins' view,
the fundamental characteristics of life are replication and evolution.  In
biological life, genes serve as the fundamental 
replicators.  In human
culture, memes are the fundamental
replicators.  Both genes and memes evolve by mutation-coated
replication and natural
selection of the fittest.  Analogous to how genes encode the essence of
biological life, Dawkins regards memes
as genetic carriers of a memotic
life form---what we know as human culture.  This theme has found
popularity in several fields 
(Waldrop 1992; Wilson 1975; Yee 1997).

Epstein's theory (1980) of evolutionary norms
goes beyond Dawkins' by 
incorporating a sociobiological
dimension.  The thrust of Epstein's hypothesis is that
human beings who abide by certain rules of conduct are more likely
to survive, reproduce, and pass on both biological genes {\it and\/} 
ideological
memes to their children.  Over generations, natural selection operating
at a {\it combined\/} biological and cultural level leads to ingrained
behavioral traits.  Epstein listed four categories of law which
he thinks has evolutionary roots: (i) prohibition against violence except
in self-defense; (ii) first possession as the root of title;
(iii) obligations of parents to their
offspring; (iv) promissory obligations.  

A cross between Dawkins' memes
and Epstein's theory of sociobiological enhancement is compelling.  In
this picture, less successful individuals and groups 
within a population must imitate
the behavior of their more successful peers in order to 
successfully compete for resources.  Accordingly, the more
above average an individual is, the more others copy his behavior.  As
a result of peer mimicry,
the population establishes and self-enforces over time
standards of normal behavior.  Normal behavior
may either be time-independent or it may cycle through a range of
behaviors.  This picture is naturally reconciled with
evolutionary games because the evolutionary process is essentially
a scenario of replication dynamics based on survival of the fittest.  Any
process which favors
iterated, merit-based growth of some subgroups at the expense of
peers---such as Darwinian evolution or
proportional group learning---can be described by
evolutionary games.

The appeal of evolutionary games is that participants 
do not have to be endowed with superhuman characteristics like
unfailing Bayesian rationality.  Even butterflies and baboons are 
qualified to play.  All that is asked
is that the parties learn by trial and error, incorporate
what they learn in future behavior, and die if they don't.

This article also raises another question, What is the connection
between norms and formal law.  The Possessor and Trade
ESSes suggest only that these ownership conventions may emerge
as social norms -- they do not draw any connection between
norms and formal law.  Indeed, Ellickson (1991) 
suggests that not only are 
fundamental norms like neighborly cooperation pervasive but
they exist {\it independently\/} {\it of\/} {\it and\/}
{\it oblivious\/} {\it to\/} 
legal standards.  Based on 
a case study of ranchers in Shasta County, California, Ellickson 
conjectures that the omnipresent threat of reputation-damaging
gossip is an effective means of enforcing social norms in
the Shasta County ranching community without the need of laws.  So
why laws?  

If law is not necessary to sustain Trade 
in Regions II or III of Figure~\ref{fig:regions}, is
the (only) role of law to guide society
into (or out of) these two Regions by twiddling cost $h$
with the exaction of penalties?

\end{document}